# A Solution to The Non-linearity of Electro-Optic Modulation Synthetic Aperture Lidar

Fuping Fang[1,2], Heng Hu[1,2], Weiming Xu[1,2], Rong Shu[1,2]

[1] Key Laboratory of space active optoelectronics, Shanghai Institute of Technical Physics, Chinese Academy of Sciences, No.500 Yutian Road, Hongkou District, Shanghai 200083, China

[2] University of Chinese Academy of Sciences, No.500 Yutian Road, Hongkou District, Shanghai 200083, China

Abstract:

Synthetic aperture laser radar has higher resolution, so requires higher modulated bandwidth. Because the data volume of chirp or pulse coding schemes is too large, it brings much pressure to data acquisition and data processing. So, we can use dechirp to reduce the amount of data. However, there is a seriously non-linear problem in phase modulation, which strictly prohibited the dechirp usage. Therefore, in order to solve the above problems, we propose a method based on Electro-Optic Modulators to achieve a large-bandwidth chirp signal. Meanwhile, give a method to resolve the seriously non-linear problems. Firstly, by properly setting the amplitude of the input signal of electro-optic modulators, the finite order signal coefficients are guaranteed to have an absolute advantage. Then, an ideal chirp signal can be obtained by filtering through an optical filter. Finally, the feasibility of the scheme is proved by theory and experiments.

Keywords: Synthetic aperture lidar, Electro-Optic Phase Modulation, Dechirp, Optical filter

Corresponding author
E-mail address: shurong@mail.sitp.ac.cn

## 1 Introduction

In recent years, the synthetic aperture laser radar (SAL) has made rapid advancement[18][6]. Similar to the synthetic aperture radar (SAR), SAL can break through the diffraction limit for long-range target imaging[1]. Compared with SAR, SAL has higher imaging resolution due to its shorter wavelength; When the detection distance exceeds 100km, SAL is the only way to provide centimeter-level imaging resolution. Therefore, SAL has extremely important values in long-distance sensing, target recognition, ground mapping, and more precise imaging[1]. In SAL, higher resolution is achieved by transmitting larger bandwidth signals such as chirp signals and pulse coding signals in the range direction[7]. In the existing papers[8][9], the LFM signal can be generated by a tunable semiconductor laser, for which it's not realistic to realize a high pulse repetition frequency needed by a real airborne or spaceborne system, whose speed is very fast. Meanwhile, because of the modulation in wavelength rather than in frequency, which causes seriously non-linear problems so its practicality is very low. Due to the development of optical communication, phase coding[10] has developed rapidly. Also, because of its good autocorrelation property, phase coding is applied to SAL as a new modulated method instead of LFM. A literature[11] reported an airborne flight experiment based on phase coding, which realized a 2cmx2.5cm two-dimensional image in the range and azimuth directions, generating a large amount of data, which brought a devastating disaster to storage and data processing. Therefore, it can be known that the resolution of SAL is mainly limited by electronics[12]. At present, there are two mainline solutions. One is increasing the Analog-to-Digital Converter's (ADC) sampling rate. According to Moore's Law, the hardware level will be updated every 1-1.5 years. Because the data volume increases according to the square rate characteristic, subsequent image processing is still full of difficulty. The other is to use dechirp processing[13], which can reduce the amount of data by 4-5 orders of magnitude.

The new proposed Electro-Optic phase Modulator (EOPM), as an external modulation, can achieve ultra-high bandwidth linear frequency modulated signal with a high pulse repetition frequency. Reference[14]proposed a method of Microwave Photonic Synthetic Aperture Radar based on mach-zehnder modulator (MZM), which achieved a 600MHZ modulation bandwidth. Reference[15]discussed dechirp technology on the basis of reference[14], so the sampling rate was reduced by four orders of magnitude. Paper[7] proposed a high-efficiency linear frequency-modulated continuous wave system based on employing an electro-optic in-phase and quadrature (I & Q) modulator with an effective bias controlling. However, the system needed real-time feedback to correct the bias of the MZM, increasing the system complexity.

In view of this, we propose a SAL system based on Electro-Optic phase Modulators realizing chirp signal, and also give a method to remove the non-linear phase caused by EOPM. Firstly, the continuous wave laser passes through an EOPM, controlled by an arbitrary waveform generator. Then after through the optical filter[16], the ideal chirp signal can be obtained. And lastly the signal is divided into two paths by a 99: 1 polarization maintaining fiber splitter. Among them, the path with 99% of power is used as the transmission signal, and the path with 1% of power is used as the reference signal. At the receiving end, the signal is dechirp-received via the free-space path. First, the signal is mixed in a 50:50 polarization beam splitter by the reference beam after passing through a pre-set delay fiber and a quarter-wave plate, which is to change the linearly polarized light into elliptically polarized light, and the received signal light. Second, the obtained I-path and Q-path signals split by above polarization beam splitter respectively pass through a half-wave plate to realize a pi-phase shift of the signal light and the reference light, and then coupled into balanced detectors. Forth, it is sampled by a dual-channel ADC. Final, after range cell migration correction (RCMC) and azimuth matching filtering, we obtain a high-resolution two-dimensional image.

## 2 Operating Principle

Figure 1 shows a based on electro-optical modulators realizing chirp signal synthetic aperture laser radar system, meanwhile, giving a method to solve non-linear problems. It consists of three parts: signal transmission, signal reception, and data processing[5]. At the transmitting end, it is consisted of a 1550nm continuous wave laser, an electro-optic modulator (EOM), an arbitrary waveform generator, an optical filter, an Erbium Doped Fiber Amplifier (EDFA), and a 99: 1 polarization maintaining fiber splitter. The continuous wave laser first passes through a LiNbO3 Electro-Optic Phase Modulator controlled by an arbitrary waveform generator. After filtering by an optical filter, the signal is divided into two paths by a 99: 1 optical beam splitter, of which 99% of the power is coupled into the transmitting lens as the transmission signal, and the 1% of the power is as the reference signal through a fixed delay optical fiber.

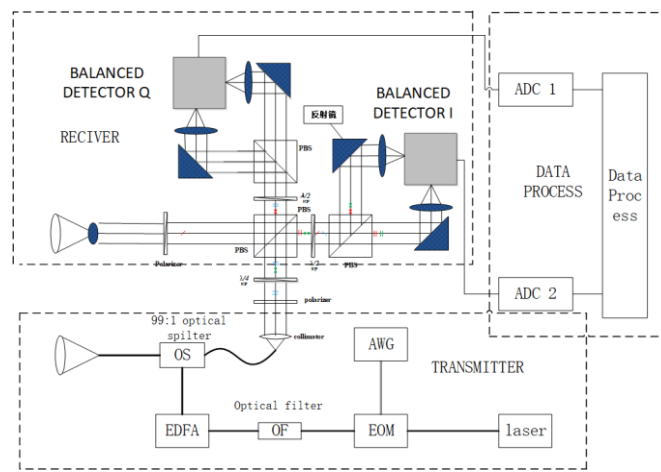

Fig1. Based on electro-optical modulators realizing chirp signal synthetic aperture laser radar system

An ideal continuous wave laser output is a single-frequency signal. However, the real laser signal is time-varying near the center frequency. Due to the effects of Gauss random frequency and Gauss random phase, its instantaneous

spectrum is wider. We refer to the commonly used laser model[17]. The laser signal output is:

$$s(t) = \cos[2\pi \int (f_c + A_F \sin(2\pi f_a))dt + 2\pi \int f_b(t)dt + \phi_c(t)] \quad (1)$$

$f_c$ is the center frequency of the laser, $f_a, A_F$ are the frequency and amplitude of the sinusoidal change of the laser signal frequency, $f_b(t) \in N(0, \delta_{fb}^2)$ is the Gaussian random frequency, $\phi_c(t) \in N(0, \delta_{\phi c}^2)$ is the Gaussian random phase, $\delta_{fb}$ and $\delta_{\phi c}$ are the standard deviations of the Gaussian random frequency and Gaussian random phase.

The arbitrary waveform generator output is:

$$g(t) = m\cos(kt^2 + 2\pi f_0 t) = m\cos(\varphi_1(t)) \quad (2)$$

The output of the laser after passing through the EOM is:

$$e(t) = \text{rect}(t/Tp) [\cos(m\cos(kt^2 + 2\pi f_0 t)) + 2\pi \int (f_c + A_F \sin(2\pi f_a))dt + 2\pi \int f_b t dt + \phi_c(t)] \quad (3)$$

Where, $\varphi_1(t)$ is the signal phase of the arbitrary waveform generator, $\varphi_2(t) = 2\pi \int (A_F \sin(2\pi f_a))dt + 2\pi \int f_b t dt + \phi_c(t)]$ is the linewidth of laser. $w_c = 2\pi f_c$ is the center angular frequency of laser signal, $f_0$ and $m$ are the amplitude and phase of output of the arbitrary waveform generator. Among them, $f_0$ is single frequency offset using to facilitate optical filtering, because of the limit of optical filters' performance, which needs a minimum interval in frequency domain. $k$ is the chirp-rate in range direction, $Tp$ is the chirp-period, and in the range direction the modulation bandwidth is $B = k * Tp$.

So, formula (3) can be expressed as:

$$e(t) = \text{rect}(t/Tp) \cos(w_c t + m\cos(\varphi_1(t)) + \varphi_2(t)) \quad (4)$$

By applying Bessel expansion[14] to Equation (4), we get:

$$e(t) = \text{rect}\left(\frac{t}{Tp}\right) \exp(j\varphi_2(t)[J_0(m) + \sum_{n=1}^{\infty}[2J_{2n}(m) \cos(w_c t + 2n * 2\pi f_0 + 2nKt^2)] \quad (5)$$

Only considering the positive frequency.

$$e(t) = \text{rect}\left(\frac{t}{Tp}\right) \exp(j\varphi_2(t)[\sum_{n=-\infty}^{\infty}[J_{2n}(m) \exp(jw_c t + 2n * 2\pi f_0 + 2nKt^2)](6)$$

$J_{2n}(*)$ is the first-kind Bessel function coefficient of the 2n-th order, $J_0(*)$ is the first-kind Bessel function coefficient of the 0th order. $m$ is the modulated index.

$$e(t) \approx \text{rect}(t/Tp)\exp(\varphi_2(t))[J_2(m)\exp(jw_c t - j2\pi f_0 - 2jkt^2)$$
$$+ J_0(m) \exp(jw_0 t)$$
$$+ J_2(m) \exp(jw_c t + j2\pi f_0 t + 2jkt^2)](7)$$

$J_2(*)$ is the first-kind Bessel coefficient of the second-order. By setting proper $m$, ensure the power is focused on Bessel coefficient of the second-order as much as possible. When $m = 1$, $e(t)$ is approximately as shown in formula (7). In Fig2.a, the blue curve is formula (7), and the red curve is formula (6). It can be seen that this approximate treatment is quick precise. In formula (7), we retain only the 0th order Bessel coefficient and the 2nd order Bessel coefficient shown in Fig.2(b), which nearly take up 99.5% of the power.

By referring to a paper[19], we can know that when the product of time and bandwidth is large enough, $s = \exp(j2\pi f_c t + j\phi_0 + j\pi kt^{\wedge}2)$. After Fourier transform, the result is approximately equal to $S(f) = \sqrt{\pi}/2k * \text{rect}((f - f_c)/kT_p)\exp(j\phi_0 + j\pi/4)$.

According to formula(6), of which the phase is $s = \sum_{n=-\infty}^{\infty}[J_{2n}(1) \exp(jw_c t + 2\pi * 2nf_0 + 2jnKt^2)]$, after by Fourier transforming, we can get the frequency domain signal as shown in Figure 2.b. The signal is separated in the frequency axis, and every partial bandwidth is equal to $kT_p$. Fig2.b(B) is the spectrum of formula(t), and Fig2.b(A) is the optical filter. As we know, the spectrum of the target signal is the part surrounded by Fig.2(A), and we can acquire the ideal signal after filtering.

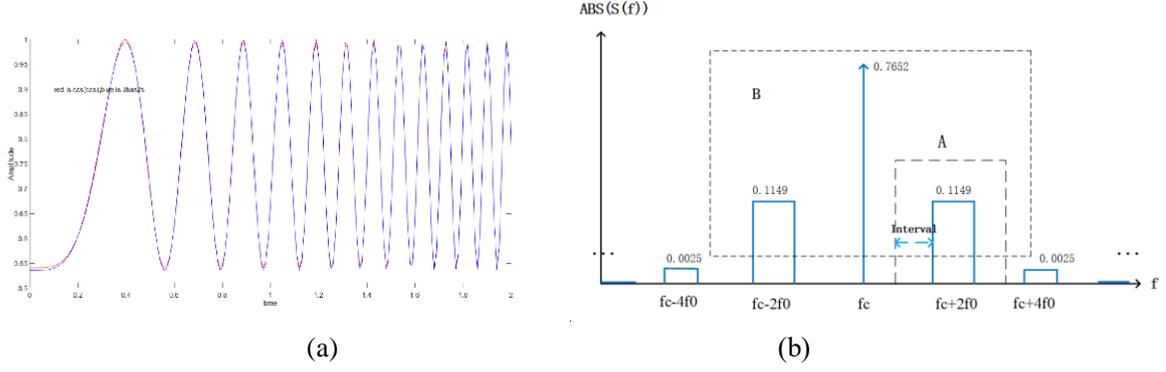

(a)                            (b)

Fig2. (a)The laser signal after through an electro-optic modulator. (b) Spectrum of the phase of electro-optic modulated signal

Then, the signal goes through an optical filter to select the -2 or +2 order Bessel coefficient as shown in Fig.2(A). Assuming the gate lets through +2 order signal, the final output is:

$$e(t) = \text{rect}(t/Tp) J_2(1) \exp(jw_c t + j2\pi f_0 t + 2jkt^2) \exp(\varphi_2(t)) \qquad (8)$$

Last, the signal is amplified by an EDFA, and the final output is the formula (9). A is the amplitude of the amplified signal.

$$e(t) = A * \text{rect}(t/Tp) \exp(jw_c t + j2\pi f_0 t + 2jkt^2) \exp(\varphi_2(t)) \qquad (9)$$

At the receiving end, the signal is dechirp-received via the free-space path. Compared with the optical fiber path, free-space has the advantages of wavelength-level stability, high integration level, and resistance to heat. The receiving system is consisted of two polarizers, two half-wave plates, a quarter-wave plate, three 50:50 polarization beam splitters, four mirrors, two balanced detectors, and a dual-channel ADC.

After passing through a predetermined delay fiber and a 1/4 wave plate, which is to change linearly polarized light into elliptically polarized light, the reference signal and the received signal light are respectively coupled from two orthogonal directions into a 50:50 PBS. After splitting, the obtained I-path and Q-path signals pass through a half-wave plate to achieve pi-phase shift between the signal light and reference light, and then they are respectively coupled into a PBS beam splitter, lastly entering a balanced detector.

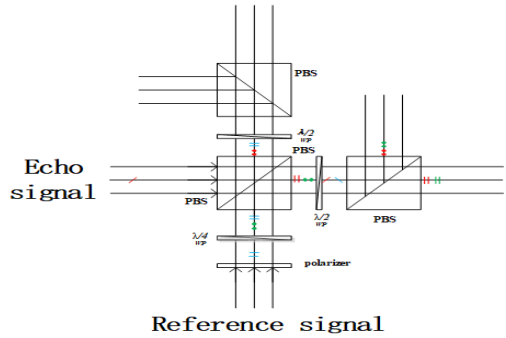

Fig3. The I&Q dechrip-receiving in free space path

Echo signal delay: $t_o = 2R_i/c$

$$s_i(t) = \sqrt{2} s e^{j\varphi_s(t-t_o)} = A_i \text{rect}((t_r - t_o)/Tp) \exp(j\varphi_2(t_r - t_o)) [\exp(j(w_c + w_0)(t_r - t_o) + 2jk(t_r - t_o)^2) \qquad (10)$$

Reference signal delay: $t_{ref} = 2R_{ref}/c$

$$s_{ref}(t) = w e^{j\varphi_w(t_r - t_{ref})} = A_{ref} \text{rect}((t_r - t_{ref})/Tp) \exp(j\varphi_2(t_r - t_{ref})) [\exp\left(j(w_c + w_0)(t_r - t_{ref}) + 2jk(t_r - t_{ref})^2\right) \qquad (11)$$

$s_i(t)$ is the echo signal, $s_i(t)$ is the reference signal, $t_r$ is the fast time. $R_i$ and $t_o$ are the real-time distance and the delay time respectively between the target and the radar; $R_{ref}$ and $t_{ref}$ are the reference distance and the reference distance delay time respectively.

The transmission Jones Matrix of the PBS is:

$$T = \begin{bmatrix} te^{j\phi_r} & 0 \\ 0 & 1 \end{bmatrix} \quad (12)$$

The reflection Jones Matrix of the PBS is:

$$R = \begin{bmatrix} 0 & 0 \\ 0 & re^{j\phi_r} \end{bmatrix} \quad (13)$$

Assuming, the laser signal is linearly polarized light, and the target does not change the laser's polarization state, so the echo signal is also the linearly polarized. In addition, in order to get the best quadrature demodulated effect, the linear polarization angle is 45 °. As mentioned above, the Jones matrix of the echo signal is:

$$S(t) = \begin{bmatrix} se^{j\varphi_s(t_r-t_o)} \\ se^{j\varphi_s(t_r-t_o)} \end{bmatrix} \quad (14)$$

The reference light is the horizontally polarized light, and its Jones matrix is:

$$S(t) = \begin{bmatrix} we^{j\varphi_w(t_r-t_{ref})} \\ 0 \end{bmatrix} \quad (15)$$

Assuming that the differences between the three PBS devices can be ignored, in most case it is also easy to satisfy.

$$R_1 = R_2 = R_3 = R$$
$$T_1 = T_2 = T_3 = R$$

The Jones matrix of a quarter-wave plate is:

$$Q = \begin{bmatrix} 1 - j\cos2\eta & -j\sin2\eta \\ -j\sin2\eta & 1 + j\cos2\eta \end{bmatrix} = \begin{bmatrix} 1 & 0 \\ -j & 1 \end{bmatrix} \quad (16)$$

After the reference signal passes through a 1/4 wave plate:

$$QW(t) = \begin{bmatrix} w(1 - j\cos\eta)e^{j\varphi_w(t_r-t_{ref})} \\ w(\sin2\eta)e^{j\varphi_w(t_r-t_{ref})+\pi/2} \end{bmatrix} = \begin{bmatrix} w(1 - j\sqrt{2}/2)e^{j\varphi_w(t_r-t_{ref})} \\ we^{j\varphi_w(t_r-t_{ref})+\pi/2} \end{bmatrix} \quad (17)$$

For a 1/4 wave plate, its main function is to transform horizontally polarized reference light into circularly polarized light, that is to say, the amplitude of the horizontal polarization component equals to the amplitude of the vertical polarization component, and the phase of the horizontal polarization component is orthogonal to the phase of the vertical polarization component. By this way, it is possible to obtain two reference signals with orthogonal polarizations and orthogonal phases using for orthogonal demodulation. By solving the above formula(17) ,we can obtain: $\eta = \frac{\pi}{4} + \frac{k\pi}{2}$, By setting $\eta = \frac{\pi}{4}$.

We can get the Jones matrix of a half-wave plate.

$$H = \begin{bmatrix} \cos2\theta & \sin2\theta \\ \sin2\theta & -\cos2\theta \end{bmatrix} = \begin{bmatrix} \sqrt{2}/2 & \sqrt{2}/2 \\ \sqrt{2}/2 & -\sqrt{2}/2 \end{bmatrix} \quad (18)$$

The reference signal and the received signal are beam split by a PBS. After splitting, the signal light and the reference light in the same optical path are polarized orthogonal. Then, the light will pass through a half-wave plate, and the output is:

$$H_1 T_1 QW(t) + H_1 RS(t) = \begin{bmatrix} rs * e^{j(\varphi_s(t_r-t_o)+\varphi_r)} \\ tw * e^{j(\varphi_w(t_r-t_{ref})+\varphi_t)} \end{bmatrix} \quad (19)$$

Secondly, the π-phase shift is generated by the horizontal and vertical polarization components of the signal and reference light with orthogonal polarization, so as to meet the phase requirements of the two balance detectors. According to the above, $\theta_1$ and $\theta_2$ are equal to $\frac{\pi}{4}$.

According to the Jones matrix of the optical path and the corresponding devices, the four paths Jones matrix can be obtained as:

$$L_1(t) = R_2 H_1 T_1 QW(t) + R_2 H_1 R_1 S(t)$$
$$L_2(t) = T_2 H_1 T_1 QW(t) + T_2 H_1 R_1 S(t)$$
$$L_3(t) = T_3 H_2 R_1 QW(t) + T_3 H_2 T_1 S(t)$$
$$L_4(t) = R_3 H_2 R_1 QW(t) + R_3 H_2 T_1 S(t) \quad (20)$$

Putting the formula(19) into the formula(20), we can be obtained:

$$l1 = \frac{\sqrt{2}}{4}[s * \exp(j(\varphi_s(t_r - t_o) + \pi)) + w * \exp(j(\varphi_s(t_r - t_o) + \pi))]$$

$$l2 = \frac{\sqrt{2}}{4}[s * \exp(j(\varphi_s(t_r - t_o))) + w * \exp(j(\varphi_w(t_r - t_{ref})))]$$

$$l3 = \frac{\sqrt{2}}{4}[s * \exp(j\varphi_s(t_r - t_o)) + w * \exp(j(\varphi_w(t_r - t_{ref}) - \frac{\pi}{2}))]$$

$$l4 = \frac{\sqrt{2}}{4}[s * exp(j\varphi_s(t_r - t_o)) + rtw * \exp(j(\varphi_s(t_r - t_o) + \frac{\pi}{2}))] \quad (21)$$

Where, assuming $t = r = \sqrt{2}/2 * (1 - \sigma)$. Choosing a suitable PBS, it is easy to ensure t=r, $\sigma$ is the power loss occurring on the PBS, which is general small, and approximately equal to 0. By properly adjust the PBS to achieve $\phi_t = \phi_r = \pi$.

The two detector outputs are:

$$I = Arect\left(\frac{t - \frac{2R_i}{c}}{T_p}\right) e^{-j\frac{4\pi}{c}\gamma\left(t - \frac{2R_{ref}}{c}\right)R_\Delta} e^{-j\frac{4\pi}{c}(f_c + f_0)R_\Delta} * e^{j\frac{4\pi\gamma}{c^2}R_\Delta^2} e^{j[\varphi_2(t_r - t_o) - j\varphi_2(t_r - t_{ref})]}$$

$$Q = Arect\left(\frac{t - \frac{2R_i}{c}}{T_p}\right) e^{j\frac{\pi}{2}} e^{-j\frac{4\pi}{c}\gamma\left(t - \frac{2R_{ref}}{c}\right)R_\Delta} * e^{-j\frac{4\pi}{c}(f_c + f_0)R_\Delta} e^{j\frac{4\pi\gamma}{c^2}R_\Delta^2} e^{j[\varphi_2(t_r - t_o) - j\varphi_2(t_r - t_{ref})]} \quad (22)$$

The laser line-width affects the azimuthal imaging resolution. [18]proposed a line width solution for random Gaussian errors. Assuming that the laser line width is small or has been corrected, the later processing method of Range Cell Migration Correction (RCMC) and azimuth compression is the same as that in[7]. By dechirp operation, the signal has achieved pulse compression. After the signal implements RCMC, the two-dimensional image can be obtained by finally implementing the azimuth compressing.

$$s(t_r, t_m) = A_0 sinc\left[\Delta f_r \left(t_r - \frac{2R_0}{c}\right)\right] sinc[\Delta f_m(t_m)] \quad (23)$$

Assuming, the target size a=1m, the frequency range after dechirp equals to $[-a/2, a/2]$, and the maximum bandwidth of the signal is $\Delta f_a = \frac{k}{c} \times \frac{1}{T} = \frac{2\Delta r}{cT}\Delta f$. It can be seen that the smaller ratio of $\Delta r/R$, the smaller sampling rate of the dechirp signal compared to the bandwidth of the original signal.

According to the Table 1, $\Delta f_a = 10GHZ$

$$\eta = \frac{a}{R_p} = \frac{a}{c * T_p} = \frac{1}{3 * 10^8 * 50 * 10^{-6}} = 0.667 * 10^{\wedge}3$$

$$\Delta f_a = \frac{k}{c} \times \frac{1}{T} = \frac{2\Delta r}{cT}\Delta f = 6.67 MHZ$$

Because the sampling rate equals to $\Delta f_s \geq 2\Delta f_a$, it can be easily seen that after dechirp processing, the sampling rate is reduced by four orders of magnitude, which is no longer difficult for electronics.

The advantage of optical fiber path is easy to operate, but the optical fiber is easily affected by vibration, temperature, etc. Free space path has the advantages of high stability, insusceptibility to thermal changes and high integration level. Normally, if a wavelength-level error occurs, the random phase error will seriously affect the phase stability, which is an inherent disadvantage of the optical fiber path. So, the free-space path is better. The structure of our solution is simple without a feedback system and without delay error. Because it is not necessary to monitor whether the phase meets the requirements in real time, it is easy to duplicate.

**3 Experiment and Simulation**

Firstly, due to the inherent non-linearity of the laser internal modulation, it is not suitable for large-bandwidth requirement. With the development of optical communication, electro-optic modulators have the characteristics of fast update rate, large modulation bandwidth, and relatively low price. However, the LiNbO3 Electro-Optic Modulator has extremely complex modulated phases. In view of this, this paper realizes the chirp signal based on an electro-optic modulator, and immediately proposes a solution to solve the nonlinear errors of Electro-Optic Modulators. secondly, by reasonably setting the amplitude of the output signal of the arbitrary waveform generator to ensure that the finite-order coefficients have absolute advantages after the signal is modulated, and then filtering by an optical filter to obtain the ideal chirp signal. Our scheme greatly expands the SAL application prospects.

Table1. The Simulation Parameters in Strip-Mode

| Parameter | Value |
| --- | --- |
| Laser Wavelength(nm)： | 1550 |
| Divergence Angle(m rad)： | 0.1 |
| Vertical Distance(km)： | 10 |
| Flying Speed(m/s)： | 50 |
| Modulated Bandwidth(GHZ)： | 5 |
| Pulse Repetition Frequency (KHZ)： | 20 |
| Azimuth Resolution (cm)： | 1.5 |
| Range Resolution(cm)： | 1.5 |

According to Fig2.b we can get the frequency domain signal. The signal is separated in the frequency axis, and every partial bandwidth is equal to $kT_p$. Fig2.b(A) is the optical filter, by which the spectrum of the target signal is the part surrounded. As we know, as long as bandwidth interval needed by the optical filter is less than $2f_0 + kT_p$, we can acquire the ideal signal after filtering, And the result shows in Fig4.

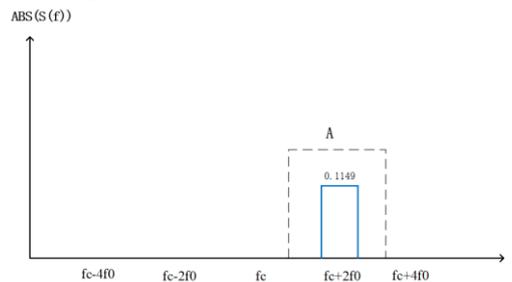

Fig4. The result after optical filtering

Assume that $B = kT_p = 5GHZ$ is the signal bandwidth of the arbitrary waveform generator(Tektronix AWG70000), $Tr = 50e - 6s$ is the duration of each modulation period and the pulse repetition frequency is $PRF = 20kHZ$. For the convenience of optical filtering, the modulation signal contains a fixed offset frequency $2f_0 = 15GHZ$. Reference[16] shows that the filter bandwidth interval of the best optical filters $\Delta\lambda = 0.2$nm.

$$\text{According to formula：} \Delta f_0 = \frac{c\Delta\lambda}{\lambda_1\lambda_2} = 26GHZ$$

Both $\lambda 1$ and $\lambda 2$ are approximately equal to 1550nm. At present, Electro-Optic Modulators (Photline MPX-LN) can easily realize a modulation signal with a bandwidth greater than or equal to 40GHZ, which is bigger than the frequency interval $(\Delta f_0 + kT_p = 31GHZ)$ required by the optical filter, so this solution is achievable.

Secondly, because the dechirp operation is used, the sampling rate is reduced by approximately 4-5 orders of magnitude, which greatly reduces the burden of hardware systems such as balanced-detectors and digital sampling systems. At the same time, due to the reduction of the sampling rate and the reduction of the sampling data, the algorithm's computation time can exponentially decrease, which brings the advantages of real-time imaging.

Lastly, we carried out three-point targets simulation, and the specific system parameters are shown in Table 1. Assume that the radar works in the side-mode[5]. Setting target 1 as the reference center, target 2 is same in azimuth direction and in range direction the offset is 3 cm. Target 3 is 2 cm away from Target 1 in azimuth direction and is same in range direction. Fig5.a is the 2-D time-domain signal after dechirp receivering. It can be seen that the two target is obviously easy distinguished, and the corresponding range resolution is better to 3cm, which is nearly equal to the theoretical value. Fig5.b is 2-D time-domain signal after RCMC. It is easy to find that the range cell migration (RCM) is very small because of the short time imaging. Fig5.c is the two-dimensional imaging map of the target. Also, we can see that, there will be great cross-influence in different targets when they are near to each other, so how to overcome the influence of the sidelobe will have greatly a lot of work to do with the improvement of resolution.

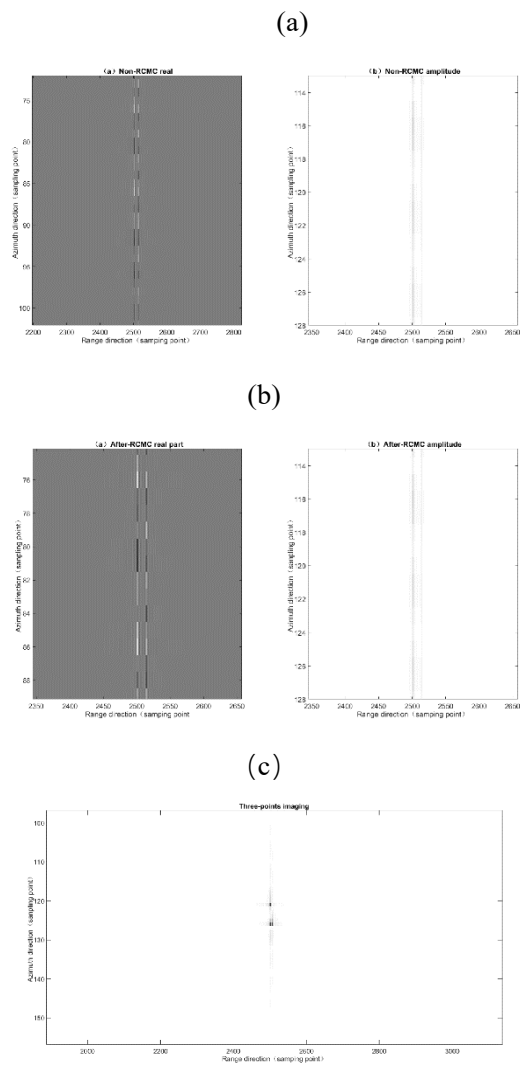

(a)

(b)

(c)

Fig5 (a) Signal after range compression. (b) Signal after RCMC. (c) Imaging results of three-point targets.

## 4 Acknowledge
The author would like to thank all the people, who have offered help.

## 5 Conclusion

Due to the requirements of higher-resolution, higher resolution corresponds to the higher modulated bandwidth in synthetic aperture laser radar. The higher the bandwidth, the more difficult it is for data acquisition and later data processing, so the resolution of SAL imaging is mainly limited by electronics. In SAL, dechirp is a good way to reduce the stress of electronics. Due to the special nature of the optical region, large-bandwidth chirp signals are not easy to implement. In view of this, this paper proposes a sheme to implement chirp signals by LiNbO3 Electro-Optic Modulators with the characteristics of large modulation bandwidth and fast update speed. As to non-linear problems, this paper presents a hardware processing method: Firstly, by appropriately setting the amplitude of the output signal of the arbitrary waveform generator to ensure that the finite-order coefficients have absolute advantages after the signal is modulated, and then through an optical filter to obtain the ideal chirp signal. Finally, by dechrip, range cell migration correction and azimuth compression, we can obtain a high-resolution two-dimensional image. The method is simple in structure and stable in work, what's more, the feasibility of the theory is proved by simulation and experiments.